%% file: my_paper.tex
\documentclass[runningheads,a4paper]{llncs}

\usepackage{amsmath}

\usepackage{amssymb}
\usepackage{graphicx}
\usepackage{url}
\usepackage{apacite}

\begin{document}

\mainmatter  

\title{Deep Music Information Dynamics}

\titlerunning{Running Title}

%
%

\author{Shlomo Dubnov} 
%
\authorrunning{Shlomo Dubnov} 

\institute{Department of Music, University of California San Diego} 

%
%

\maketitle

\begin{abstract}
    Music comprises of a set of complex simultaneous events organized in time. In this paper we introduce a novel framework that we call Deep Musical Information Dynamics, which combines two parallel streams - a low rate latent representation stream that is assumed to capture the dynamics of a thought process contrasted with a higher rate information dynamics derived from the musical data itself. Motivated by rate-distortion theories of human cognition we propose a framework for exploring possible relations between imaginary anticipations existing in the listener's mind and information dynamics of the musical surface itself. This model is demonstrated for the case of symbolic (MIDI) data, as accounting for acoustic surface would require many more layers to capture  instrument properties and performance expressive inflections. The mathematical framework is based on variational encoding that first establishes a high rate representation of the musical observations, which is then reduced using a bit-allocation method into a parallel low rate data stream. The combined loss considered here includes both the information rate in terms of time evolution for each stream, and the fidelity of encoding measured in terms of mutual information between the high and low rate representations.  
    In the simulations presented in the paper we are able to juxtapose aspects of latent/imaginary surprisal versus surprisal of the music surface in a manner that is quantifiable and computationally tractable. The set of computational tools is discussed in the paper, suggesting that a trade off between compression and prediction are an important factor in the analysis and design of time-based music generative models.
\end{abstract}
\input{paper_contents}

\section*{Acknowledgment}
I would like to thank the reviewers for the detailed and insightful comments. This work is partially supported by Cygames, Inc.
\bibliography{my_paper}
\bibliographystyle{apacite}

\end{document}

%% file: paper_contents.tex
\section{Introduction}
Music Information Dynamics is a field in music analysis that is inspired by theories of musical anticipation \cite{meyer56emotion}\cite{huron06sweet}, which deals with quantifying the amount of information passing over time between past and future in musical signal \cite{dubnov06spectral},\cite{Abdallah}. Music Information Dynamics can be estimated in terms of Information Rate, which is defined as mutual information between past and future of a musical signal. Generative models that maximize information rate were shown to provide good results in machine improvisation systems 
\cite{pasquier2017introduction}. Since music is constantly changing, the ability to capture structure in time depends on the way similarity is computed over time.
The underlying motivation in proposing a deep information model is to acknowledge the fact that imagination, both for the composer, improviser and the listener, is playing an important and possibly even a crucial role in experiencing and creation of music. Music generation and listening are active processes that involve simultaneous processing of the incoming musical information in order to extract salient features, while at the same time predicting  the evolution of those features over time, an aspect that builds anticipations and allows creation of surprise, validation or violation of expectation and building of tensions and resolutions in a musical narrative.

\section{Deep Information Dynamics}
\begin{figure}[!htb] 
\centering
\includegraphics[width=65mm]{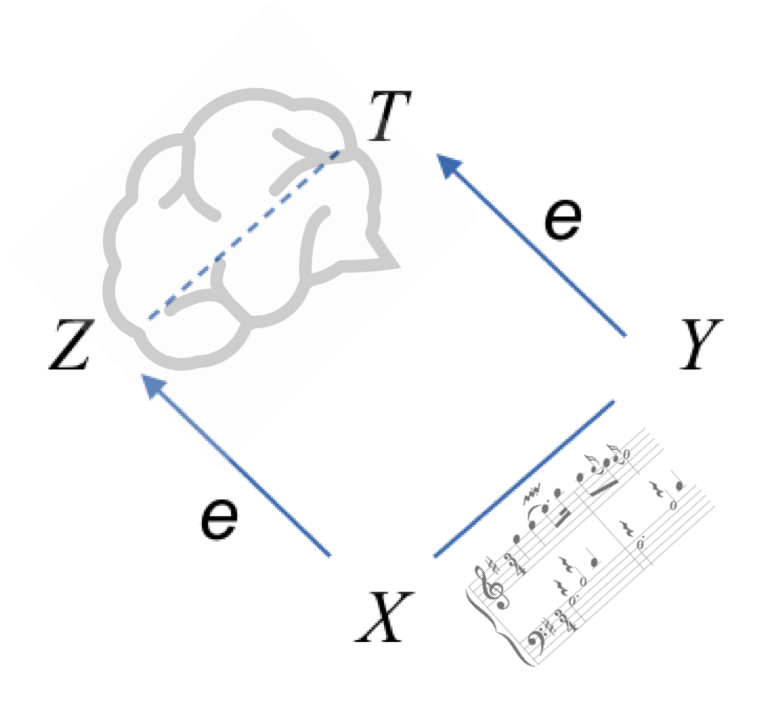}
\caption{Graph of the model variable statistical dependencies. The letter "e" represents an encoding that will be parametrized according to different complexity through changing the bit-rate between the encoder and decoder in VAE.}\label{fig:Graph-Model}
\label{fig:1}
\end{figure}

In order to allow quantitative approach to analysis of what's going on in the musical mind, we propose an information theoretic model for the relation between four factors: the signal past $X$, the signal present $Y$, and their internal or mental representation in terms of past and present latent variables $Z$ and $T$, respectively. This highly simplified model assumes a set of Markov chain relations, as shown in Figure \ref{fig:1}, between the past of the signal $X$ that is encoded into a latent representation $Z$,  the future of the signal $Y$ that depends on its past $X$, and its approximation by a latent representation $T$ that is predicted from past latent representation $Z$. 
Using Markov relation $Z-X-Y$, we see that $X$ "shields" the future $Y$ from the past latent $Z$, meaning that once the past musical surface $X$ is considered, there is no additional information or statistical dependency between the next musical surface $Y$ and past internal state $Z$. According to this rule we can try to formulate a mathematical expression for the goals underlying the learning process of such a musicing system \cite{Musicing}. Our expression for the optimization goal comprises of a trade off between simplicity of representation and its prediction ability. Accordingly, we are looking for representation that is minimizing the discrepancy, or statistical distortion, measured by Kullback–Leibler divergence $D_{KL}$, between signal prediction of $Y$ using complete information about the past $X$, versus its prediction capability by using a simplified encoding of the past $Z$. Using $I(X,Y) = H(Y)-H(Y|X)$ to denote mutual information, with $H(\cdot)$ as the entropy, the overall quality of such error averaged over all possible encoding pairs $X,Z$ of the musical surface and its latent code becomes
\begin{align}
   \langle & D_{KL}(p(Y|X)||p(Y|Z)) \rangle_{p(X,Z)} = I(X,Y|Z) \nonumber \\ 
   & = I(X,Y)-I(Z,Y)  ~. 
    \label{eq:PredKL}
\end{align}
Since $I(X,Y)$ are independent of $p(Z|X)$, minimizing the error between the true conditional probability of the future on its past $p(Y|X)$, compared to probability of the future conditioned on the latent representation $p(Y|Z)$, requires minimization of $-I(Z,Y)$, or maximizing the mutual information between the encoding of the past and the signal future.  

Using Markov relations between the past and present latent-variables according to diagram shown in Fig.\ref{fig:1}, we express $I(Z,Y) = I(Z,T)-I(Z,T|Y)$. This shows that the ability of predicting the future of the musical surface $Y$ from the latent (mental) representation of the past $Z$, measured by their mutual information, comprises of a  difference between the information dynamics of the latent representation, measured in terms of the mutual information involved in imagining the next latent representation $T$ from past latent states $Z$, and the residual or redundant information between these latent states $Z$ and $T$ once the actual musical surface $Y$ is revealed or heard by the listener. In other words, the amount of information between latent past and latent future states is being reduced once the actual next instance of the musical surface is revealed to the listener, and this difference between imagined music future "in the brain" versus actually the surprise in hearing the next musical event amount to the quality factor in equation \ref{eq:PredKL} that represents the the ability of the system to predict the next musical surface from its past internal state.

If we assume that the latent representation fully captures the surface, or in other words, if $T=Y$, then full knowledge about the musical surface is already contained in the imaginary latent states sequence, resulting in zero residual information $I(Z,T|Y)=0$. In such case we may ignore the right side of the equation, which creates an exceptional situation where there is no need to actually listen to the music and the best experience is achieved simply by imaginary prediction. In case when the mental representation is not perfect, a non-zero surprisal factor $I(Z,T|Y) = I(Z,T) - I(Z,Y)$ allows for musical tension to emerge during listening, or be deliberately inserted by the creator during composition or improvisation.

\subsection{Surface Representation and Information Dynamics}
In our information theoretical approach one needs to know the probability distribution of the relevant variables in order to compute the appropriate information measures. Since we do not know the true probability dynamics of the musical surface $p(X,Y)$, we will substitute it by a variational approximation by encoding it into latent codes $Z$ and $T$ using a Variational Auto-Encoder (VAE). It was shown that minimization of $I(X,Z)$ under distortion constraint $D(X,Z)$ is equivalent to learning a VAE representation through minimizing of Evidence Lower Bound (ELBO)\cite{BrokenELBO}. 
Combining ELBO and the prediction objective, gives us a combination of latent encoding quality and temporal information  
\begin{equation}
    \mathcal{L} = I(X,Z) + \beta \langle D(X, Z) \rangle - \gamma (I(Z,T)-I(Z,T|Y)) ~.
    \label{eq:DeepRD}
\end{equation}
It should be noted that in the above expression there are separate variables referring to past $X$ and $Z$, and future $T,Y$. In the following we will first train a neural network using  past data so as to minimize the $X,Z$ part of the loss, and then manipulate $Z,T$ by bit-rate limited encoding. In the following we will assume that $X=Z_{full-rate}$ and $Y = T_{full-rate}$, and compare it to lower bit-rate encoding. 

\subsection{Bit-rate limited Encoding}
\begin{figure}
    \centering
    \includegraphics[width=4cm, height=6cm]{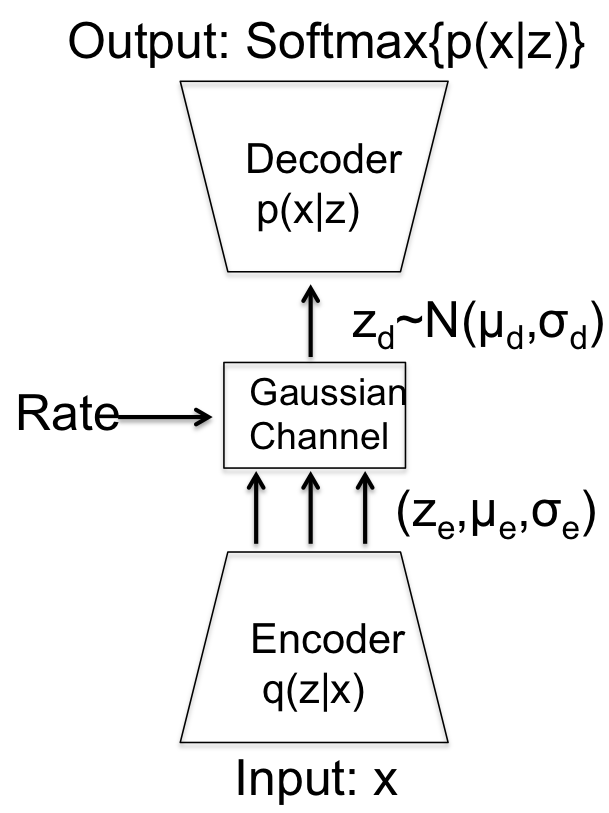}
     \caption{Noisy channel between encoder and decoder}
    \label{fig:AE_noisy_channel}
\end{figure}

Using a noisy channel between encoder and decoder we are able to control the complexity of the encoding using bit-allocation. The rate-distortion theory quantifies the trade off between the amount of information between two variables, measured by their mutual information, and the distortion or error between them. This theory is a basis for lossy compression, where less bits need to be transmitted for lower quality signals. According to this theory, for a simple Gaussian information source of variance $\sigma^2$, the rate $R$ and given distortion level $D$ is given by
\begin{equation}
    R(D) = 
    \begin{cases}
        \frac{1}{2} log_2 \frac{\sigma^2}{D}, & \text{if $0 \leq D \leq \sigma^2$}\\
        0, & \text{if $D > \sigma^2$}.
    \end{cases}
\end{equation}
One can see that for distortions above variance level, no bits need to be transmitted. The bit-allocation algorithm uses the above equation to allocate different amount of bits to multiple variables, which in our case are the latent variables of the VAE encoder. Starting from the highest variance, or strongest variable, it iteratively allocates bits in optimal manner, until the bit pool is exhausted. Then lower resolution variables are then used to generate new outputs through a decoder. Schematic representation of the channel is given by Figure \ref{fig:AE_noisy_channel}.
Encoding for a lower bit-rate is given by the following optimal channel \cite{Berger}
\begin{align}
     Q(z_d|z_e) & = Normal(\mu_d,\sigma_d^2) \\
    \mu_d & = z_e + 2^{-2R}(\mu_e - z_e) \\
    \sigma^2_d & = 2^{-4R}(2^{2R}-1)\sigma^2_e
\end{align}

For each latent variable, the equation tells us the mean and variance of the decoder's conditional probability. One can see that a channels with zero rate will transmit deterministic  mean value of that element, while channels with infinite rate will transmit the input values with zero added noise. 

\subsection{Estimation of Information Dynamics using VMO}
The computation of Information Dynamics is done in terms of Information Rate, IR=$I(X,Y)$, which is a measure of mutual information between past and present in a time series. This requires a predictive model that can capture the joint information between past and present by learning from examples. Deep models such as RNN can be used to model time sequences, and predictive information measures can be implemented using estimators of mutual information between the last hidden variable in RNN that summarizes the whole sequence, and the predicted variable. 
One of the difficulties in using RNN is the limited history or poor modeling of long sequences. 

Variable Markov Oracle (VMO) \cite{Wang2014a} is a method based on the Factor Oracle (FO) string matching algorithm that initially quantizes a signal $x_1^T = x_1,x_2,\dots, x_T$, into a symbolic sequence $s_1^T = s_1,s_2,\dots,s_T$, over a finite alphabet $s \in S$, with $X=x_{past}=x_1^{T-1}$ and $Y=x_{present}=x_T$. 
IR is estimated by applying a string compression method $C$ and using an approximation $I(X,Y) = H(Y)-H(Y|X) \approx C(Y) - C(Y|X)$, where we substitute entropy $H$ with compression $C$, with $C(Y) = log_2(|S|)$ being the number of bits required for individual symbol encoding over alphabet $S$, and $C(Y|X)$ being a block-wise encoding that recursively points to repeated sub-sequences, such as in the Lempel-Ziv or Compror string compression algorithms \cite{Wang2015}.
In the following we apply VMO  to estimate the information dynamics of the latent representation $I(Z,T)$ at different bit-rates.

\subsection{Estimation of the Predictive Encoding Quality}
In the VAE training step we  minimized $I(X,Z)$ (and $I(Y,T)$ as well) that gives an optimal instantaneous representation. Assuming $Y = T_{full-rate}$, and $Z = Z_{limited-rate}$, we use MINE (Mutual Information Neural Estimation) \cite{MINE} as a method for estimating $I(Z,Y)$. The network used in the experiments comprises of two parallel networks with shared weights, one receiving ordered pairs of $Z,Y$ and the other receiving a pair of $Z$ with a shuffled version of $Y$, both mapped though two fully connected layers with 30 hidden states with a dropout layer of $0.3$ and eventually mapped to a single output. 
The training was done until approximate convergence, as shown in the results section. 

\begin{figure}[!htb] 
    \includegraphics[width=\columnwidth]{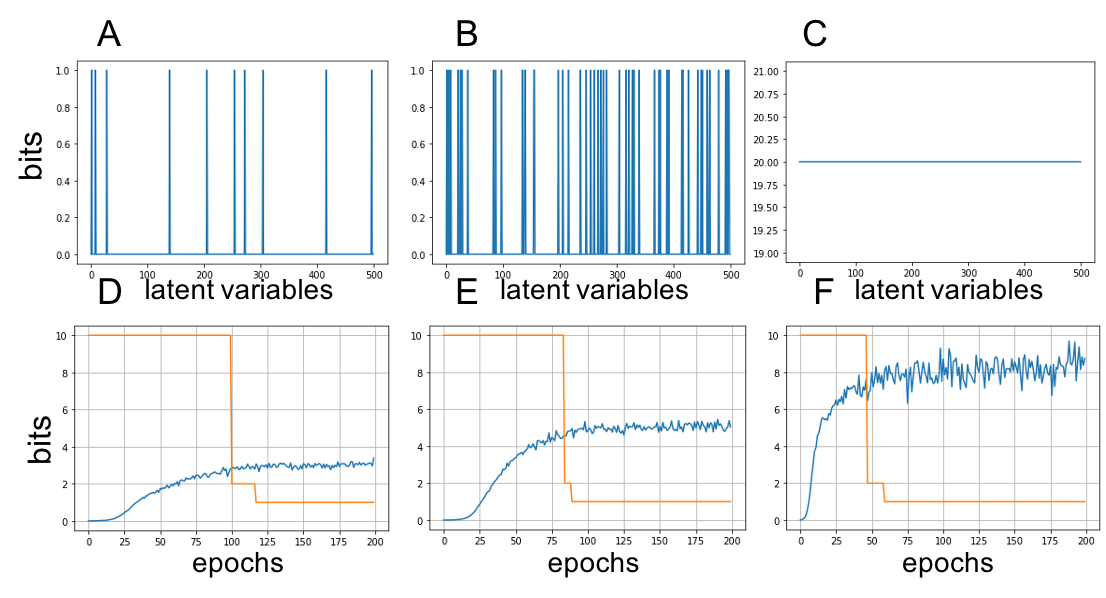}
    \caption{Estimation of the predictive quality of bit-limited representation. Sub-figures A,B,C show the bit-rate allocation at rates 10, 50, and 10000, respectively. The x-axis corresponds to the 500 latent state variables, with portions showing no-bit allocation basically not being transmitted or accounted for in the latent representation. Sub-figures D,E,F show the MINE estimate as function of the training epochs, for these rates.}
    \label{fig:3}
\end{figure}

\section{Experimental Results}
In order to demonstrate the utility of the proposed model for music analysis, we show the results of deep information rate analysis of a single musical piece. The concept of surprisal $I(Z,T|Y) = I(Z,T) - I(Z,Y)$ was defined as the difference between the ability to imagine the next latent state and the ability to predict the next musical surface. From pure mathematical perspective both factors require averaging over the complete data. In practice, VMO provides instantaneous measure of information rate since we have access to compression rate at every time step of the time series.
The MINE method averages over all data pairs $Z,Y$, outputting a single number. In future work we plan to train a predictor in order to compute an instantaneous signal (surface) prediction error, so that both factors of surprisal could be considered in time. 

The experiments reported below were conducted on a MIDI file of the Prelude and Fugue No. 2 in C Minor, BWV 847, by J.S.Bach. The full rate representation was obtained by training a VAE encoder on set of MIDI files from LABROSA\footnote{\url{https://labrosa.ee.columbia.edu/projects/piano/}}. The VAE that was used for the encoding had a single fully connected hidden layer of size 500, trained with ELBO loss function with Cross-Entropy reconstruction and KL reparameterisation loss using an Adam optimizer, which are the standard settings for VAE. 
Figure \ref{fig:3} shows the predictive quality, measured as mutual information $I(Z,Y)$ between a bit-reduced latent representation $Z$ and the music signal $Y$ one bar into the future (we remind that $Y$ is represented by $T_{full-rate}$ of the VAE encoder). The bit-allocation regime for 500 latent states is shown at the top row. The x-axis of the top row are indices of the latent vectors, and the y-axis is the number of bits. The bottom row shows the MINE optimization process that converges to an estimate of the mutual information $I(Z,Y)$ when it reaches the plateau after about 100 epochs. The results show that for rate 10 the amount of predictive information is around 3 bit, at rate 50 around 5 bits and at full (10000 bit) rate, it is between 8 and 9 bits. Such result can be expected since more complex (higher bit-rate) latent representation $Z$ carries more information about the future of the signal $Y$.

\begin{figure}[!htb] 
    \includegraphics[width=\columnwidth]{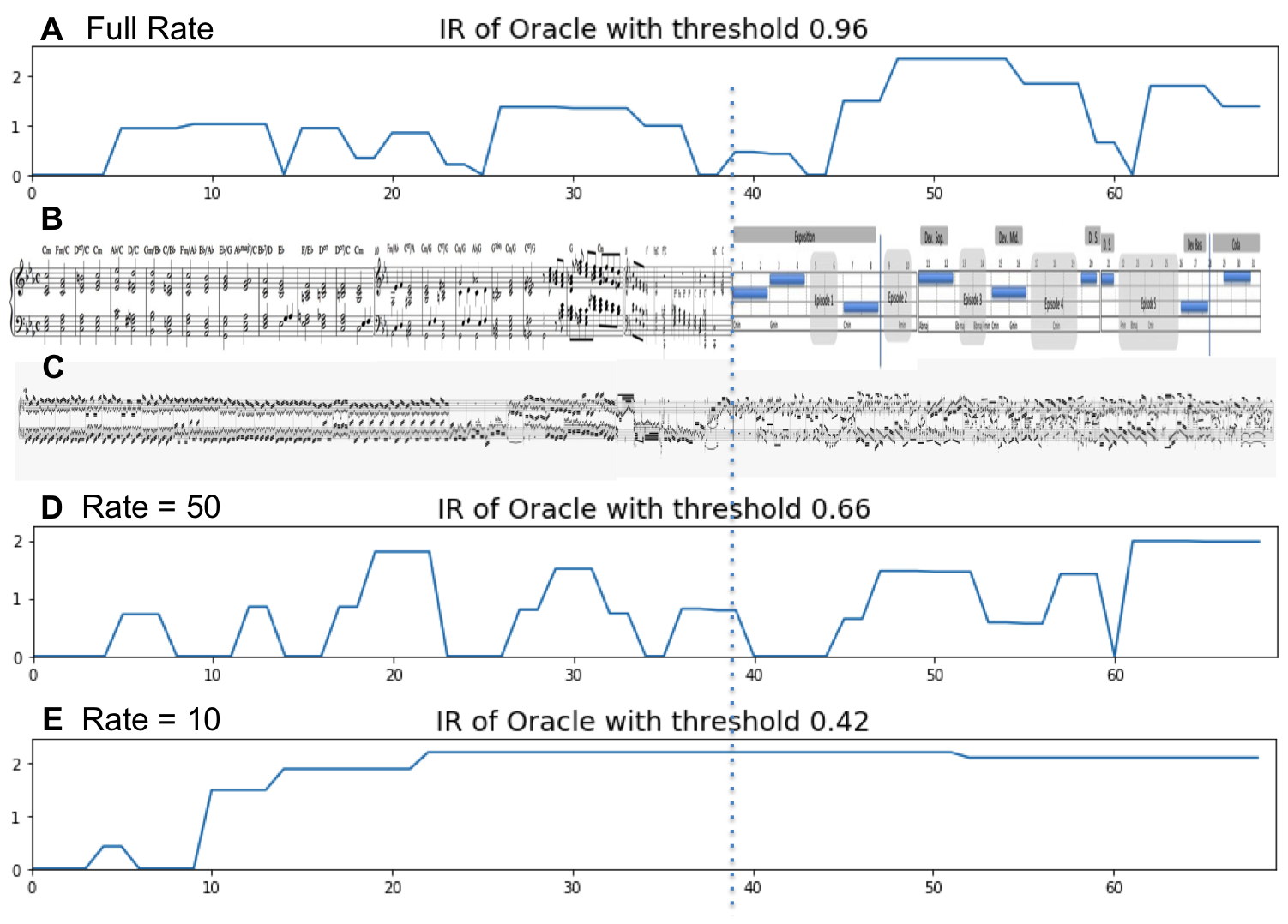}
    \caption{Analysis of Information Rate using VMO at A:full-rate, D: rate 50, E: rate 10. Sub-figure B shows harmonic analysis of the Prelude and Thematic analysis of the Fugue, Sub-figure C shows the musical score.}
    \label{fig:2}
\end{figure}

Figure \ref{fig:2} shows the information dynamics IR=$I(Z,T)$ of the latent representations itself at different representation complexity levels. This analysis is meant to capture the imaginary expectation of music based on the dynamics of the latent representation alone. A bit-allocation algorithm was used to reduce the representation complexity of the VAE encoding to the desired bit-rate. For musical reference we provide plots of harmonic and thematic analysis of the piece\footnote{\url{http://bachwelltemperedclavier.org/pf-c-minor.html}} in sub-figure B, and score rendering in sub-figure C.

 


\section{Discussion and Summary}
In this paper we presented a work in progress for developing a framework for modeling of musical surprisal, formulated in terms of information theoretical relations between full-rate (high-fidelity) encoding of the musical data, and a lower complexity latent encoding that models mental or imaginary musical representations. This formalizes the notion of musical anticipation that were proposed by various researchers in terms of information dynamics and representation learning, taking into account the limited capacity of cognitive processing and the trade off in fidelity of its representation of sensory input. It is evident from the experiments that lowering the bit-rate of the encoding has a dramatic effect on the information dynamics of the latent representation. 
Considering information dynamics of latent states as expectations formed by our imagination, the points where the expectations differ at different bit-rates are assumed to carry creative or experiential significance. In the experimental results one can see that transition to new materials in bars 25-27 causes a drop in Information Rate. Also the development of thematic material in the Figure starting around bars 44-45 increases IR in both full and 50 rate\footnote{The units of bit-rate reduction are total bits per measure}. 
For rate of 10 we see that except for a short initial period where all materials are still new (low IR), the rest of the music is perceived as one long repetition (high IR). It should be noted that these results may be also due to the nature of the piece itself and the quality of the encodes. Additional analysis on multiple pieces and different encoding and predictive architectures are anticipated in the future.

The motivation for this type of modeling comes from the cognitive idea that musical creativity and musical perception obey a trade-off between abstraction or simplified representation of music that captures more salient or structural aspects of music, and perceptual sensibility to the musical surface that is abundant in detail. To the best of our knowledge, no such conceptual or computational framework had been previously offered. 
